\documentclass[final]{rmaa}

\usepackage{paralist}

\usepackage{psfrag,color}

\usepackage[latin1]{inputenc}
\usepackage{natbib}
\bibpunct{(}{)}{;}{a}{}{,}
\newcommand{\magdot}[1]{\ensuremath{^{\rm m}\!\!#1\,}}
\newcommand{\hst}{{\it HST}}
\newcommand{\ergl}{~\ensuremath{\rm erg~s^{-1}\/ }}

\newcommand{\kms}{~\ensuremath{\rm km~s^{-1} \/}}

\newcommand{\oiii}{[\ion{O}{3}]}
\newcommand{\oii}{[\ion{O}{2}]}
\newcommand{\sii}{[\ion{S}{2}]}
\newcommand{\nii}{[\ion{N}{2}]}
\newcommand{\heii}{\ion{He}{2}}

\newcommand{\Mpc}{~\ensuremath{\rm  Mpc \/}} 
\newcommand{\pc}{~\ensuremath{\rm  pc \/}}
\newcommand{\Myr}{~\ensuremath{\rm  Myr \/}}
\newcommand{\cmc}{~\ensuremath{\rm  cm^{-3} \/}}
\title{Kinematics of the Nebular Complex MH9/10/11 Associated with HoIX~X-1}

\author{P.~Abolmasov and A.~V.~Moiseev
\affil{Special Astrophysical Observatory RAS, Russia}
}

\shortauthor{Abolmasov \& Moiseev}
\shorttitle{Kinematics of MH9/10/11}

\fulladdresses{
\item      Special Astrophysical Observatory,
    Nizhnij Arkhyz,
    Zelenchukskij region,
    Karachai-Cirkassian Republic,
    Russia 369167
}

\listofauthors{P.~Abolmasov \& A.~V.~Moiseev}
\indexauthor{Abolmasov,~P.}
\indexauthor{Moiseev,~A.~V.}

\abstract{
We report the results of our observations of the nebular complex
 MH9/10/11, associated with the ULX HoIX~X-1, with scanning
 Fabry-P\'erot Interferometer. 
Two regions differing by their kinematics and line ratios may be 
distinguished, roughly corresponding to the bubble nebula MH9/10 and
fainter HII-region MH11. 
For MH9/10 we find the expansion rate of $20\div 70 \kms$ that is
different for the approaching and receding parts.
MH11 is characterised by very low velocity dispersion
 ($\lesssim 15\kms$) and nearly constant line-of-sight velocities.
Properties of MH11 may be explained by photoionization
of gas with hydrogen density of
$ \sim 0.2\,\cmc$. Luminosity required for that should be of the
order of $10^{39}$\ergl. Similar power source is required to explain
the expansion rate of MH9/10.
Modelling results also indicate that oxygen abundance in MH11 is
about solar.
}

\addkeyword{ISM: bubbles}
\addkeyword{ISM: individuals: MH9/10/11}
\addkeyword{ISM: kinematics and dynamics}
\addkeyword{X-ray: Individuals: HoIX~X-1}
\begin{document}
% Typeset article header
\maketitle

\section{Introduction}\label{sec:intro}

The nature of Ultraluminous X-ray sources, or ULXs, is first addressed
in the work of \citet{fabbiano88}.
These objects were a subject of intense study for the past 20 years and remain
one of the unresolved problems in astrophysics \citep{roberts_review}.
Optical observations show that many of these sources are surrounded by
large-scale (from tens to hundreds of parsecs) nebulae.
We review the properties of some ULX Nebulae (ULXNe) in \citet{list}.

Recent works on ULX environment \citep{ngc7331,ramsey} show that many
of these objects are associated with young (several million years) stellar population,
supporting the hypothesis that ULXs are a certain class of accreting
binaries with high-mass donor stars. Young SNRs and X-ray bright SNe
are excluded from ULXs by definition
though their properties in X-rays may be similar, see discussion in
\citet{fabbiano88} and references therein. Very often ULXs are found in
merging and starburst galaxies.

HoIX is a post-starburst tidal dwarf galaxy lacking old stellar
population. We adopt here a distance of 3.6\Mpc\ measured by
\citet{m81group}. According to \citet{RC3}, line-of-sight velocities of the galaxy
are equal to $46\pm 6\, \kms$ for neutral hydrogen (HI 21cm) and $119\pm
60\, \kms$ for the stellar component.

\citet{MH94} present a survey of all the bright HII-regions in M81 group
dwarf galaxies including HoIX in a narrow-band filter sensitive to
H$\alpha$ and \nii$\lambda$6583.
The three brightest HII-regions detected in HoIX form a single extended structure:
a bright shell (numbers 9 and 10, according to \citet{MH94}) with some fainter nebulosity (MH11) to the
southeast. Spatial dimensions of MH9/10 are $300\pc \times 400$\pc.
Subsequent work by \citet{miller} identifies the bubble with M81~X-9, or HoIX~X-1,
that is one of the oldest known ULXs \citep{fabbiano88}.
The coordinates of the X-ray source as measured by Chandra % from {\it Chandra} data
are $\alpha = 09^h 57^m 53^s\!\!.25$, $\delta =  +69\arcdeg\, 03\arcmin\,
48\farcs{}3$ (J2000). With accuracy of about $ 0\farcs{}5$ the X-ray source coincides with
a relatively bright star with $V \sim 23\magdot{\,}$ \citep{ramsey}.

The X-ray source and its environment were extensively studied during
the last two decades.
It was shown that optical emission lines in the spectrum of MH9/10 are
broadened \citep{ramsey} suggesting that the nebula is powered by shock
waves.
Optical spectra were acquired with low spectral resolution
\citep{miller,list} revealing some new features such as \heii$\lambda$4686 emission
from the vicinity of the X-ray source.
\hst\ observations \citep{ramsey} show that
 the X-ray source coincides with a young stellar association.
Isochrone fitting points to an age in the range $4\div 6$\Myr.
\citet{ramsey} detect 5 stars in the mass range $12\div 20\,M_\odot$
implying that the
total mass of the association is of the order of $ 10^3 M_\odot$.
The authors argue that supernova explosions and stellar winds are short in
explaining the observed luminosity and the size of the bubble.

\citet{hoix_pakull} report that in high-ionization \oiii$\lambda$5007
emission line MH11 is about as bright as MH9/10. 
This points to somewhat different physical conditions in MH11 that
may be a consequence of different ionization and heating mechanisms. 

Our kinematical study is aimed to acquire more information
about both the shell and the high-ionization part of
the nebular complex. 
In the next section we describe our observations
with scanning Fabry-P\'erot Interferometer.
The main results are given in section~\ref{sec:res}.
We analyse the results for MH9/10 in section~\ref{sec:mh910} and for
MH11 in sections~\ref{sec:mh11} and~\ref{sec:cloudy}, the latter
devoted to photoionization modelling. 
The results are discussed in section~\ref{sec:disc}.

\section{Observations}\label{sec:obs}

Our observations were carried out on January 15/16, 2008 in the prime focus
of the Russian Special Astrophysical Observatory 6m telescope with the
SCORPIO multi-mode focal reducer \citep{scorpio}.
We used a scanning Fabry-P\'erot Interferometer (FPI) providing
spectral resolution  $30\div35$\kms. 
The object was observed in two emission lines:  $\sii\lambda$6717 (total
exposure $160\rm\,s \times 36$ spectral channels) and  
$\oiii\lambda$5007 (total
exposure $180\rm\,s \times 36$ spectral channels). 
The free spectral range was $13.7$ and $7.7$\AA, correspondingly. 
Seeing was around $2\div 2.5\arcsec$ during the observations. 
The detector was EEV~42-40 $2048\times2048$ CCD operated with binning
 $4\times 4$ to reduce the readout time. 
The spatial scale is $0\farcs{}7$ per pixel.

Reduction was performed in IDL environment using {\tt ifpwid} software
designed by one of us (A.V.M.). Data reduction algorithms 
are described by \citet{ifpred} and \citet{ifpred2}.
Line profile parameters were determined by fitting with Voigt
functions of fixed Lorentzian widths
($30$\kms\ for \oiii$\lambda$5007 and $34$\kms\ for \sii$\lambda$6717). 
Instrumental profile was measured using the spectra of
He-Ne-Ar calibration lamp.
Voigt fitting procedure allows to measure line widths even when they
are less than the instrumental profile width \citep{ifpred2}. 
Profiles were fitted only in the pixels where flux
exceeded 18~ADU (corresponding to $S/N \sim 3$).
All the line-of-sight velocities presented here are heliocentric.

\section{Results}\label{sec:res}

\begin{figure*}
  \includegraphics[width=0.9\textwidth]{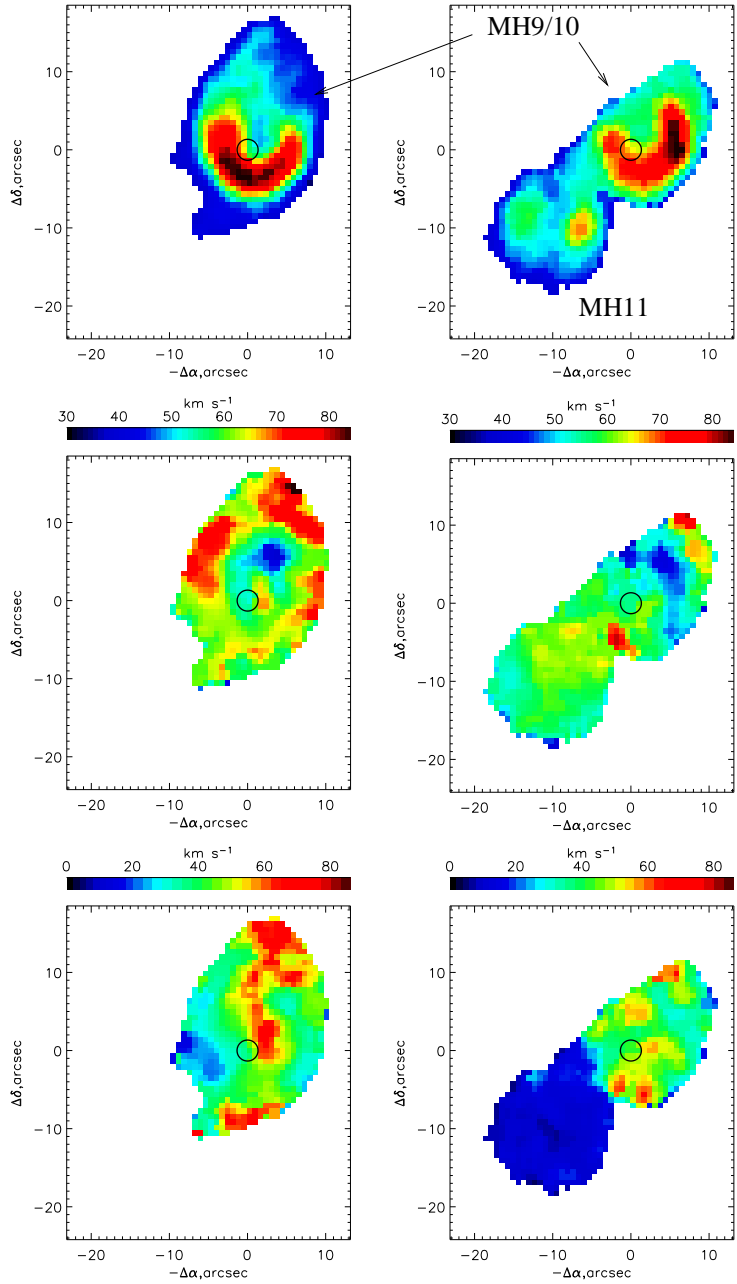}
  \caption{
From top to bottom: intensity, velocity and velocity dispersion maps in the two emission
lines (\sii$\lambda$6717 to the left, \oiii$\lambda$5007 to the right).
Velocity and velocity dispersion scales are given above the
corresponding pictures. X-ray source is shown by a 1\arcsec\ radius
circle, coordinates are given relative to the ULX. 
}
\label{fig:allmaps}
\end{figure*}

Line intensity, line-of-sight velocity and velocity dispersion maps are presented in
figure~\ref{fig:allmaps}. 
All the maps where smoothed by a $3\times 3$ median filter.
It may be seen that the two parts of the nebula have similar size but
vastly different kinematics and line ratios. Velocity dispersion is
$\lesssim 15\kms$ for MH11 but generally exceeds $ 30\kms$ for  MH9/10.
MH11 is definitely seen in the \oiii\ line but not in \sii$\lambda$6717. As we will
see below, %% OK!
 \oiii$\lambda$5007 / H$\beta$ flux ratio differs by a
factor of $\sim 10$ for the two nebulae.
\oiii$\lambda$5007 flux from MH11 is $0.65 \pm 0.1$ of that from
MH9/10 (the uncertainty is due to the uncertain boundary between the
two regions). 

Information about line profiles from different parts of the nebular
complex is summarized in table~\ref{tab:velties}. Generally, fitting
with Voigt profiles was used with appropriate Lorentzian widths.
We select several regions of interest shown by black (central) and
white (offset regions) rectangles in figure~\ref{fig:profiles}.
In order to measure the expansion rate of the nebula we select a
rectangular region (30 pixels, $2\arcsec \times 7\arcsec$) near the center
of the bubble where expansion should result in mostly line-of-sight
motions.  \sii\ line profile in the central region is asymmetric and we fit it with a double
gaussian. Parameters of both components are given in table~\ref{tab:velties}. 
The intensities of the two components are $0.28\pm 0.05$ and $0.72\pm
0.05$ of the total line intensity.

 Line profiles are extracted also from three offset regions of the 
 same size located to the West, to the East and to the South from the
 X-ray source (shown by white dotted lines is figure
~\ref{fig:profiles} and denoted as W, E and S in table~\ref{tab:velties}).

We also integrate line profiles over the pixels with low velocity
dispersion ($D \le 20\kms$) in \oiii$\lambda$5007 that belong mostly
 to MH11. Hereafter we refer to them as the profiles extracted from MH11. 
Some pixels at the outer rim of MH9/10 have very narrow unshifted
\oiii\ line profiles (see \oiii\ line dispersion map in
figure~\ref{fig:allmaps}) 
therefore we suggest that the nature of the emitting gas is the same
with that in MH11.
In figure~\ref{fig:profiles} the profiles of both lines extracted from
the central region are presented together with the profile of \oiii$\lambda$5007
from MH11. The mean line-of sight velocity of MH11 is $V_{MH11} =
58\pm 2\kms$, that is close to the line-of-sight velocities of both 
the neutral gas and the stellar component of HoIX. 
It is also identical within the uncertainties with the line-of-sight velocities
 of the offset regions in the \sii\ line. 

We did not make any flux calibrations. However, we used line
luminosities for MH9/10 from \citet{list} corrected for Galactic
absorption. Total line luminosities for MH11 are derived using MH9/10
as calibrator. 
The luminosities and sizes of the two parts of the
nebula are given in table~\ref{tab:charvals} together with {\it
  Cloudy} modelling results (see section~\ref{sec:cloudy}). 
Line luminosities for MH9/10 in table~\ref{tab:charvals} are given
according to \citet{MH94} and \citet{list} with flux corrections for
the Galactic extinction of $A_V = 0\magdot{.}26$.
Line luminosities for MH11 are estimated using fluxes measured 
from the two nebulae in our FPI data. \sii\ flux is estimated by
integrating the spatial elements where the \oiii\ emission is detected
and its velocity dispersion is $\le 20\kms$.

\section{MH9/10}\label{sec:mh910}

Kinematical properties of the nebula (such as asymmetric
\sii$\lambda$6717 profile in the central region and velocity shift
between the central and peripheral parts) may be explained by its
asymmetric expansion.
\sii\ velocity of the peripheral regions is
consistent with the velocity of the dynamically quiet gas in MH11
hence we consider the systemic velocity equal to 60\kms.
Two-gaussian fit of the \sii$\lambda$6717 line profile shows two
velocity components.
Line-of-sight velocities of the components are $37\pm 1\kms$ and $131\pm 3\kms$. 
Velocity shifts with respect to the systemic velocity of the bubble are $-23$ and
$71\kms$, respectively, implying that the expansion is anisotropic.
The approaching part of the shell is about two times brighter, and its
velocity is more than three times closer to the systemic velocity.

\begin{table*}\centering
\caption{Line velocities and velocity dispersions in different parts
  of the nebular complex. E, W and S correspond to three spatially
  distinct regions at the bubble periphery (see text). \sii\ line profile in
  the central region is fitted with two gaussian components with velocity
  dispersion pegged at 100\kms. 
} \label{tab:velties}
% old version:
%  \setlength{\tabcolsep}{1.2\tabcolsep} \tablecols{10}
\small
%  \begin{changemargin}{0cm}{0cm}
    \setlength{\tabcolsep}{0.4\tabcolsep} \tablecols{10}
\begin{tabular}{p{3cm} @{\hspace{4\tabcolsep}} p{7.75cm}} % line -- region -- {Vr -- sigmaV}
      \toprule
  line & \begin{tabular}{p{3cm} @{\hspace{4\tabcolsep}} p{2cm} p{2cm}} region &  $V_r$,\kms  & $\sigma_V$,\kms \\ \end{tabular} \\
      \midrule
% \hline
\oiii$\lambda$5007  &
\begin{tabular}{p{3cm} @{\hspace{4\tabcolsep}} p{2cm}p{2cm}}
  MH9/10 (center) & 50$\pm$2  &  46$\pm$2 \\  % Voigt 30km/s
  MH9/10E  & 54$\pm$2 & 36$\pm$2  \\ % Voigt 30km/s
  MH9/10W & 45$\pm$1 & 36$\pm$1  \\  % Voigt 30km/s
  MH9/10S & 56$\pm$1 & 45$\pm$2  \\ % Voigt 30km/s
  MH11 & 59$\pm$1 &  12$\pm$1 \\ % Voigt 30km/s
\end{tabular}\\
\noalign{\medskip}
\cmidrule(r){2-2}
\noalign{\medskip}
\sii$\lambda$6717   &
\begin{tabular}{p{3cm}  @{\hspace{2\tabcolsep}} p{2cm}p{2cm}}
   MH9/10 (center) &
  \multicolumn{2}{l}{
 \begin{tabular}{p{2cm}p{2cm}}
 37$\pm$3  &  100 \\ % Multigaus
  131$\pm$8  &  100 \\ % Multigaus
 \end{tabular}}\\
\noalign{\medskip}
 MH9/10E  &\hspace{2\tabcolsep}  62$\pm$1 & \hspace{1\tabcolsep} 34$\pm$1  \\ % Voigt 34 km/s
 MH9/10W &\hspace{2\tabcolsep}  59$\pm$1 &\hspace{1\tabcolsep}  37$\pm$1 \\ % Voigt 34 km/s
MH9/10S &\hspace{2\tabcolsep}  58$\pm$1 &\hspace{1\tabcolsep}  40$\pm$1  \\ % Voigt 34 km/s
 MH11 &\hspace{2\tabcolsep}  60$\pm$1 & \hspace{1\tabcolsep}  24$\pm$1 \\ % Voigt 34 km/s
 \end{tabular}
\end{tabular}
%\end{changemargin}
\normalsize
\end{table*}

Total power of the shock wave may be estimated using expressions
from \citet{DoSutI}.
 Assuming the shell spherical
 and integrating expression (3.3) from \citet{DoSutI} over a spherical
 shock front expanding with constant velocity one obtains: 

\begin{equation}\label{E:power}
L_{tot} = 7 \times 10^{39} R_{150}^2 V_{50}^3 n_{10} \ergl,
\end{equation}
\noindent
where $n_{10}$, $R_{150}$ and $V_{50}$ are correspondingly the
preshock hydrogen density (in $10$\cmc\ units), the shell radius in
$150\,\pc$ units and the shock velocity in
$50\,\kms$ units. The formula is expected to be valid for radiative $20\div
100\,\kms$ interstellar shocks and does not account for precursor
emission. $50\,\kms$ value is taken as the arithmetic mean of the measured
expansion velocities. 
If one assumes a constant energy influx responsible for powering the
nebula, a power of about $10^{39}\div 10^{40}$\ergl\ is needed
(depending on the ambient gas density)
similar to the apparent luminosity of the X-ray source.

Balmer line luminosities are consistent with the expansion
velocity estimates made above, if the mean density of the unshocked
material is about $5\,\cmc$.
Observed H$\beta$ luminosity of MH9/10 is $(2.73 \pm 0.13) \times
10^{37}$\ergl\ (or higher if additional extinction is present). This
value may be compared with H$\beta$ luminosity calculated using
expression (3.4) provided by \citet{DoSutI}:

 \begin{equation}\label{E:shop}
L(H\beta) = 3.8 \times 10^{37} V_{50}^{2.41} R_{150}^2 n_{10} \ergl.
\end{equation}

A mean pre-shock density of about $5\div 10\,\cmc$ is needed to explain the 
luminosity in H$\beta$. 
Multiple shock fronts and anisotropic expansion velocity may
be responsible for this rather high effective pre-shock density
value.
Emergent emission line flux from a unit shock front surface area scales as
$F \propto V^{2.41} n$ with the shock velocity and preshock
density. Applying this scaling to the flux from the central region
points to a $\sim 20$ times higher pre-shock density for the
approaching section of the bubble.

\begin{figure*}
  \includegraphics[width=\columnwidth]{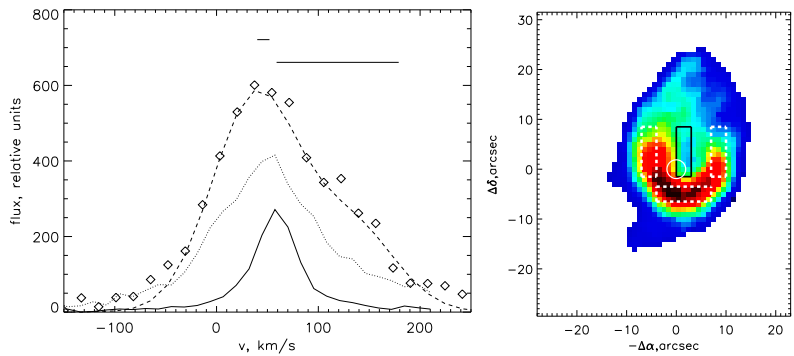}
  \caption{
Left: profiles of \oiii$\lambda$5007 integrated over the pixels with low
  velocity dispersion (solid line, downscaled by a factor of 10) and over the central area of the bubble
  (dotted). \sii$\lambda$6717 line profile from the same central
  region is shown by diamonds,
  dashed curve respresents the two-gaussian fit.
Horisontal bars correspond to velocity estimates for stars
  (lower bar) and HI (upper). Right: rectangular region (solid line) defined as the
  central part of the bubble. Three offset regions are shown by white
  dotted lines.
}
\label{fig:profiles}
\end{figure*}

In figure~\ref{fig:profiles} it may be seen that the \oiii\ line has
a narrow unshifted component present even in the central parts of the
bubble. Since $50\kms$ shock waves are uncapable for creating
precursors we conclude that the unshifted component is
emitted either by the warm gas inside the bubble
or by a photoionized region similar to MH11 on the line of
sight to the bubble.

\section{MH11}\label{sec:mh11}

We confirm the early results of \citet{hoix_pakull} that
\oiii$\lambda$5007 emission is extremely bright in MH11.
From the MPFS observations reported in \citet{list} we know the total luminosity
(corrected for Galactic absorption only) of the shell in the \oiii\ line,
$L(\oiii\lambda 5007) = (3.97 \pm 0.12) \times 10^{37}$\ergl.
Luminosity of MH11 in the same line is therefore $2.6 \times 10^{37}\ergl$.
Because H$\alpha$+\nii$\lambda$6583 luminosities of MH11 and MH9/10
differ by a factor of $10$ \citep{MH94}, H$\beta$ luminosity of MH11 should be
close to $ 2.7 \times 10^{36}\ergl$. \oiii$\lambda$5007
/ H$\beta$ flux ratio is $ \sim 10$ or higher (if \nii$\lambda$6583 / H$\alpha$ ratio
is enhanced in the high-excitation nebula). All the line luminosity
estimates are given in table~\ref{tab:charvals}.

\begin{table*}\centering
\caption{Line luminosities and approximate sizes of the two parts of the nebular complex.
In the last column spatial sizes (radius for MH9/10, diameter for
MH11) are given. 
Uncertainties in radii reflect deviations from circularity. 
Last two rows correspond to {\it Cloudy} model nebulae.
}\label{tab:charvals}
\small
\begin{changemargin}{-1.5cm}{-1.5cm}
\begin{tabular}{l @{\hspace{7\tabcolsep}}  c c c c
  c  @{\hspace{4\tabcolsep}}  c}
%p{1.3cm}p{1.8cm}p{1.8cm}p{1.8cm}p{1.8cm}p{1.8cm}} % nebula -- oiii lum -- sii lum -- size
\toprule
& \multicolumn{4}{c}{Line luminosities, $10^{37}$\ergl} && \\
\cmidrule{2-5}
nebula &  H$\beta$ & H$\alpha$+\nii$\lambda$6583 & \oiii$\lambda$5007 & \sii$\lambda$6717 && $R, \pc$ \\
\midrule
MH9/10  & 2.73$\pm$0.13 & 11.8$\pm$0.5   &  3.97$\pm$0.12 &  4.3$\pm$0.14  && 150$\pm$50 \\
MH11    & $\sim$0.27    & 1.12$\pm$0.05  &  2.6$\pm$0.4    &  0.2$\pm$0.02 && 200$\pm$50 \\
\noalign{\medskip}
{\it Cloudy} ({\scriptsize Z=Z$_\odot$ })&  0.26  & 1.6 & 2.4           &  0.4
  && 180 \\ 
\noalign{\medskip}
{\it Cloudy} \par ({\scriptsize Z=0.2Z$_\odot$})& 0.84   &  3.0           & 4.0            &  0.3
&& 250 \\ 
\end{tabular}
\end{changemargin}
\normalsize
\end{table*}

Quiet kinematics and high \oiii$\lambda$5007 / H$\beta$ flux
ratio favour photoionization as the main energy source in
MH11. 
Balmer lines are likely to be recombination lines.
H$\beta$ luminosity is determined by the number of ionizing quanta while
the \oiii\ doublet is collisionally
excited and is enhanced effectively by additional
heating. X-ray and harder EUV radiation may be that additional energy
source.
Low recombination line luminosity of a large nebula may be a
consequence of the low recombining gas density. Let us consider
MH11 a sphere of a radius $R = 100\,\pc$. 
Assuming the gas completely homogeneous one may estimate hydrogen density as:

\begin{equation}\label{E:ne_anal}
n_H \simeq \left(\frac{L(H\beta)}{E(H\beta) \alpha_{eff}(H\beta) V}\right)^{1/2},
\end{equation}
\noindent
where $V = \frac{4\pi}{3} R^3$ is the volume of the nebula and $\alpha_{eff}(H\beta)
\sim (1\div 2)\times 10^{-14} \rm cm^3 s^{-1}\/$ is the effective recombination coefficient for
H$\beta$ at $(1\div 3)\times 10^4\,\rm K\/$ in the low-density limit
\citep{osterbrock}. $E(H\beta)$ is the energy of an H$\beta$ photon.
Finally one may estimate the mean hydrogen density in MH11 as:

\begin{equation}\label{E:ne_fig}
n_H \simeq 0.22 \left(\frac{L(H\beta)}{2.7\times
 10^{36}\ergl}\right)^{1/2} 
\left(\frac{R}{100\pc}\right)^{-3/2}.
\end{equation}

\section{Photoionization Modelling}\label{sec:cloudy}

In order to better understand the physics of MH11 and to estimate the
parameters of the ionizing source
we calculated two {\it Cloudy} \citep{cloudy} photoionization models. 
Version 07.02.00 of the code was used.
We considered the nebula to be a spherical sector with a covering factor
  $0.3$ in order to reproduce the offset position of the nebula with
  respect to the X-ray source. 
Geometry was considered open (using closed geometry alters the output
  parameters by $10\div 15\%$).
Gas was irradiated by an EUV blackbody source with variable
temperature and luminosity.
We used {\tt optimize } command to find the optimal solution predicting
\oiii$\lambda$5007 / H$\beta$ flux ratio and H$\beta$ luminosity
closest to the observed values.
Hydrogen density was taken equal to $0.2\,\cmc$. 
Two abundance sets were used, solar ({\tt HII region}
  abundance set) and $1/5$ solar ({\tt HII region} abundance set with
  all the heavy-element abundances reduced by a factor of 5).

\begin{figure*}
  \includegraphics[width=\columnwidth]{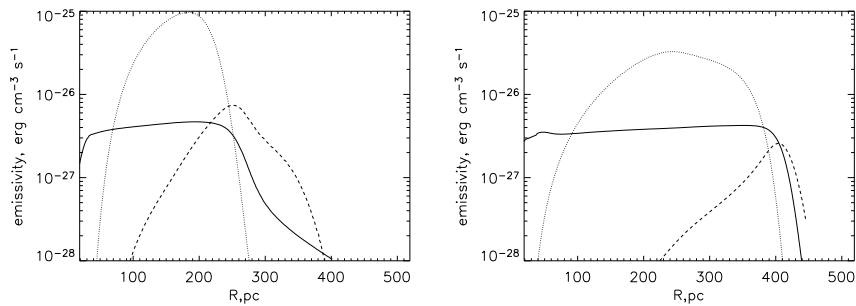}
  \caption{
{\it Cloudy} model line emissivities as functions of the radial
coordinate.
Left panel shows the results for solar, right panel for 0.2 solar metallicity. 
 Solid, dashed and dotted lines correspond to H$\beta$,
\sii$\lambda$6717 and \oiii$\lambda$5007, respectively. 
}
\label{fig:clouds}
\end{figure*}

The best-fit parameters are
$T_{BB} = 3\times 10^5\,K$ and $L = 1.9 \times 10^{39}\ergl$ for the
solar-metallicity, and $T_{BB} = 1.2\times 10^5\,K$ and $L = 3 \times
10^{39}\ergl$ for the subsolar metallicity model. 
Actually only 30\% of these luminosities are used in calculations
because of the covering factor.
In table~\ref{tab:charvals} we present line luminosities and sizes
obtained for the best-fit models. Radii are calculated as the radii
of the regions emitting \oiii$\lambda$5007. In lower ionization lines
such as \sii$\lambda$6717 the nebula is expected
to be about two times larger, therefore we may underestimate the actual
luminosity in the \sii\ line. Line emissivities for the two models are
shown in figure~\ref{fig:clouds}. Note that the source is located
outside the nebula and the observed diameter of the nebula should be
compared with the model radii. 

The best-fit solar-metallicity model predicts $L(H\beta) \simeq 3 \times 10^{36}\ergl$ and
\oiii$\lambda$5007 / H$\beta$ $\simeq 9 $ in reasonable agreement with
the observational data. It also predicts that the size of
the model nebula should be close to $ 200\,\pc$\ not taking into account
the faint low-excitation nebulosity present at larger radii due to
X-ray radiation. 
The {\it Cloudy} model also predicts bright
\oii$\lambda$3727 emission (about as bright as the
\oiii$\lambda$5007 line) and relatively bright low-excitation
lines such as \sii$\lambda$6717,6731 doublet with luminosities comparable to Balmer
line luminosities. 

The best-fit subsolar metallicity model predicts \oiii$\lambda$5007 / H$\beta$ $\simeq
4$ and severely overestimates the H$\beta$ luminosity of the nebula. 
We conclude that subsolar metallicity models have difficulties
in reproducing the observed \oiii$\lambda$5007 / H$\beta$
ratio possibly indicating that oxygen abundance is around solar for the
nebula rather than $0.1\div 0.2$ solar reported by \citet{miller}.
It is probably even higher because both models overestimate
H$\alpha$ + \nii$\lambda$6583 luminosity. 
Certainly, more thoroughful investigation involving larger number
of emissions is needed.

%%%%%%%%%%%%%%%%%%%%%%%%%%%%%%%%%%%%%%%%%%%%%%%%%%%%%%%%%%%%%%%

\section{Discussion}\label{sec:disc}

\subsection{Photoionizing Source}

Existence of ULX nebulae supports the idea that ULXs (or at least some of them) are
supercritically accreting binaries similar to SS433
\citep{katz86}. That analogy allows two energy sources of comparable
power to exist that may be responsible for powering the nebulae: jet activity
(jet power is of the order $10^{39}\ergl$ in the case of SS433) and
photoionizing radiation from the X-ray source. Both are likely to
produce HII-regions elongated in the disc/jet symmetry axis direction.

Observational properties of MH11 are consistent with photoionization
and heating by a powerful EUV and X-ray source. High EUV luminosities
(comparable with the apparent isotropic luminosities in X-rays)
of ULXs are supported both by theory \citep{poutanen} and by observations
\citep{mf16_pasj}.
Therefore, HII-regions similar to MH11 should be common for ULXs. 
Indeed, there are sources like M101P98 \citep{list,kuntz}
surrounderd by extended HII-regions with high \oiii$\lambda$5007 /
H$\beta$ ratios as well as bubble nebulae overlapped by diffuse
structures seen in \oiii$\lambda$5007 / H$\beta$ intensity maps as ``ionisation
cones'' \citep{rgww}. It is possible that in many cases
high-ionization photoionized nebulae are
masked by ULX bubbles that have about an order of magnitude higher luminosities in
Balmer lines.

\subsection{Dynamical Properties of the Bubble}

\citet{ramsey} proved that MH9/10 could not be produced by
SNe and stellar winds from the parent association of the ULX.
There is also evidence that ULX bubbles are produced by continuous
power injection by wind or jet activity rather than by instantaneous powerful
explosions \citep{hoix_pakull,funasdalen}.
If one assumes a continuous source of power that heats the gas inside
a wind-blown cavern, expansion law established by \citet{avedisova} 
(see also \citet{castor}) for pressure-dominated bubbles may be used:

\begin{equation}\label{E:predomR}
R = 70 n_{10}^{-1/5} L_{39}^{1/5} t_6^{3/5} \pc,
\end{equation}
\noindent
\begin{equation}\label{E:predomV}
V = 40 n_{10}^{-1/5} L_{39}^{1/5} t_6^{-2/5} \kms.
\end{equation}

Here $L_{39}$ is the power of the energy source in $10^{39}\ergl$
units, $n_{10}$ is the preshock density in $10\,\cmc$ and $t_6$ is
the bubble age in million years. 
These formulae may be reversed to find the kinematical age and
the power of the energy source:

\begin{equation}\label{E:predomT}
t = 7 \times 10^5\,R_{150} V_{50}^{-1} \,yr,
\end{equation}
\noindent
\begin{equation}\label{E:predomL}
L = 2.8 \times 10^{39} R_{150}^2 V_{50}^3 n_{10} \ergl.
\end{equation}

Dynamical age $\sim 1\Myr$ is typical for ULX bubbles
\citep{PaMir} but higher values were never found supposing the
lifetimes of ULXs are of the order of $\sim$1\Myr.

\subsection{Underlying Density Gradient}

Narrow-band images of MH9/10 with higher spatial resolution reveal
fine details at the outer boundary of the bubble \citep{hoix_grise}
and faint filamentary nebulosity extending to about twice the mean radius of the shell.
The complex structure of the bubble is probably connected with a high
ambient density gradient. The mean preshock
density was probably one or two orders of magnitude higher then the
density of the gas in MH11.

Ambient density
gradients often lead to blow-out structures and multiple shock
fronts \citep{maclow_blowouts}.
In this scope, it is tempting to consider MH11 a blow-out with an
invisible outer boundary. The observed gas was ionized by shock waves
and is recombining without any additional energy source. % filled with recombining gas.
Recombination time for the rarefied warm gas is long
enough: $t_{rec} \sim 1 / n_e \alpha \sim 1\,\Myr$. 
The strongest argument against
the hypothesis is the quiet kinematics of MH11. If a shock wave
propagates in a non-homogeneous medium its velocity varies
roughly as $v\propto n^{-1/2}$ \citep{mckee}. 
Disturbed gas behind the shock
front should have both high velocity dispersion due to turbulent
motions and also a velocity component in the shock propagation
direction.

\begin{figure*}
  \includegraphics[width=\columnwidth]{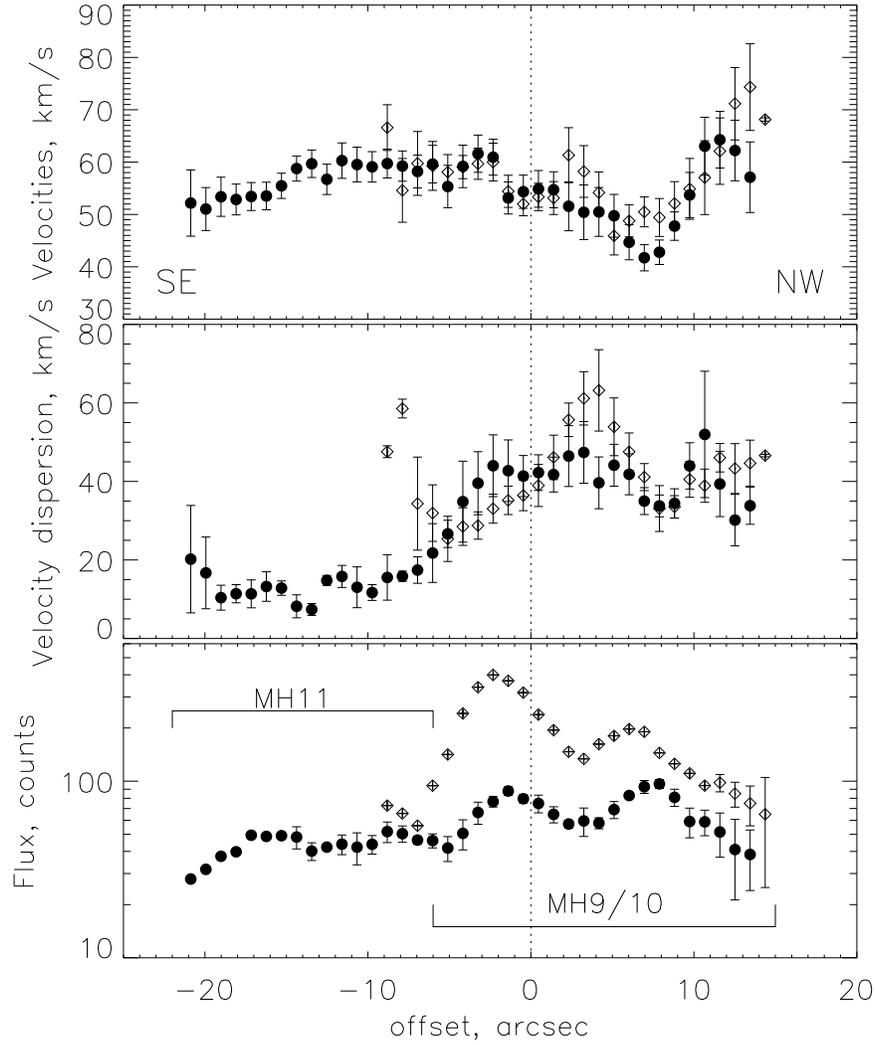}
  \caption{
Cross-section of the line parameter maps by an artificial slit
($3\farcs{}5$ wide) passing through
the X-ray source at a positional angle $131\arcdeg$. Filled circles
represent \oiii\ line profile parameters and diamonds represent those
of the \sii\ emission. Offset along the slit is given along the
abscissa, zero value corresponds to the X-ray source position.
}
\label{fig:slices}
\end{figure*}

We see however neither high velocity dispersion (such as tens \kms\ or
 higher) that should
inevitably appear if a fast shock was responsible for ionizing the
gas nor any strong line-of-sight velocity gradients. Line centroid
shifts smoothly by less than 10\kms\ towards the outer rim of the
nebula. 
This shift may appear if the HII-region is expanding due to internal
 pressure of the warm ionized gas \citep{osterbrock}. The expansion velocity is close to
 the speed of sound in the ionized gas, that is of the order of $ 10\kms$.
In figure~\ref{fig:slices} we show the behavior of
line profile parameters along the line crossing the central parts of
both nebulae at a positional angle of $131\arcdeg$.

Our observations are the first to state that velocity gradients are
not an essential part of the dynamics of ULX nebulae. The peak of the
\oiii\ line changes its line-of-sight velocity by less than 10\kms\
for MH11.
That is an important clue indicating that on large
scales the momentum injected in the ISM is low. ULX nebulae are likely to be
powered by radiation and/or relativistic jets that transport negligible
amounts of momentum for a given mechanical luminosity. 

\section{Conclusions}\label{sec:conc}

Observations with scanning FPI reveal new details about the extended
nebular complex associated with HoIX~X-1.
We measured the expansion rate of MH9/10 and find it consistent with
the velocity estimates from H$\beta$ luminosity. 
However, the expansion appears to be anisotropic.
Approaching and receding parts of the bubble have line-of-sight
velocities shifted by $-23$ and $71$\kms\ with respect to the systemic
velocity of $60$\kms. 
Complex structure of the shell probably originates from density gradients
that are definitely present in the ISM in HoIX and may be connected to
the parent association of the ULX.
The dynamical age inferred from the kinematical data is $t\simeq
0.7\,\Myr$.
Mechanical luminosity required is $L \simeq 3\times
10^{39}\ergl$, that is comparable to the X-ray luminosity of the source. 
The effective value of the pre-shocked density is $5\div 10\,\cmc$.

We show that the observational properties of MH11, namely its high
\oiii$\lambda$5007 / H$\beta$ ratio ($\sim 10$), size ($\sim
200\,\pc$) and H$\beta$ luminosity ($\sim 3 \times 10^{36}$\ergl) may be
explained by a hard EUV source ionizing low-density gas with $n_H \simeq
0.2\,\cmc$. This is the best evidence for an EUV source associated
with a ULX for today. Solar oxygen abundance value explains the
observational properties of MH11 better than 1/5~solar.
The EUV source is well reproduced by a black body with $T \sim
(1\div 2) \times 10^5\,K$ and isotropic luminosity $L \sim (1\div 3)\times 10^{39}\ergl$. 
We suggest that further observations
are needed in order to deside on the abundances and the ionization balance in MH11. 

\bigskip 

This work is  based on the observational data obtained with the
6-m telescope of the Special Astrophysical Observatory of the
Russian Academy of Sciences funded by the Ministry of Science of
the Russian Federation (registration number 01-43). 
We would also like to thank the anonymous referee for his/her very
useful remarks and suggestions and S.~Pavluchenko for his help with
English.

\end{document}